# Enhanced metamagnetic shape memory effect in Heusler-type $Ni_{37}Co_{11}Mn_{43}Sn_9$ polycrystalline ferromagnetic shape memory alloy

Sudip Kumar Sarkar[1, 4*], P. D. Babu[2], V. K. Sharma[3, 4], S. D. Kaushik[2], Srikanta Goswami[2,5] and M. A. Manekar[3]

[1]Materials Science Division, Bhabha Atomic Research Centre, Mumbai - 400085, India

[2]UGC-DAE Consortium for Scientific Research, Mumbai Centre, BARC, Mumbai - 400085, India

[3]Accelerator Physics and Synchrotrons Utilization Division, Raja Ramanna Centre for Advanced Technology, Indore 452013, India

[4]Homi Bhabha National Institute, Training School Complex, Anushakti Nagar, Mumbai - 400 094, India

[5]FZU - Institute of Physics of the Czech Academy of Sciences, Na Slovance 1999 / 2, 182 21 Prague 8, Czechia.

[*]E-mail address:s.sudip.iitg@gmail.com

**Abstract:** Polycrystalline Ni-Co-Mn-Sn based ferromagnetic shape memory alloys (FSMAs) show promise as actuator materials, but their practical application involving magnetic field induced strain (MFIS) is often limited by three factors: the requirement for high magnetic fields (> 5 T), martensitic transition temperature away from room temperature, and limited recovery of pre-strain applied to the martensite phase. Current work investigates the martensitic transition (MT) and shape memory effect under the application of magnetic field for bulk polycrystalline $Ni_{37}Co_{11}Mn_{43}Sn_9$ alloy. The outcome of the study reveals a metamagnetic transition from the martensitic phase to the austenitic phase at a low field of 2.8 T at 300 K which results 0.25% spontaneous MFIS. Interestingly, 1.3% pre-strained specimen registers a 100% recovery with the application of magnetic field of 4.5 T. Furthermore, the pre-strained specimen exhibited a two-way shape memory effect between a strain value of 1.0% to 1.55% during the field loading and unloading sequences. Notably, this study also demonstrates, to the best of our knowledge , for the first time, that the spontaneous strain and pre-strain add together. This finding paves the way for achieving a giant MFIS by pre-straining a Ni-Mn-Sn/In class of FSMAs which shows large spontaneous MFIS.

# 1. Introduction:

The discovery of large magnetic field-induced strains in Ferromagnetic Shape Memory Alloys (FSMAs) has sparked significant interest among researchers worldwide due to their potential as actuator materials [1–7]. Large non-contact deformation from magnetic field induced phase transformation (MFIPT) in FSMAs puts them in advantageous position over the conventional thermo-elastic NiTi based shape memory alloys [8–10]. Magnetic Field Induced Strain (MFIS) in FSMAs can arise either from magnetic field driven martensitic variant orientation or through MFIPT [11]. In general, these alloys undergo phase transformation from high temperature austenite phase with high crystallographic symmetry to low temperature martensite phase with lower symmetry on lowering the temperature [12]. The strain originating from the change in lattice dimensions during martensitic transition is accommodated in the martensite through formation of self-accommodated twin structure [12]. On application of magnetic field, martensite variants can rotate themselves along the direction of the field, provided the magneto-crystalline anisotropy energy (MCAE) is high enough. During this process, the variant which has a higher MCAE grows at an expense of other variants, leading to a macroscopic shape change [3,13]. Conventional Ni-Mn-Ga based FSMAs generate a reversible strain as high as 10% by virtue of the former mechanism but the associated actuation stress is quite low, typically around 3-4 MPa [3]. This limitation stems from the MCAE, which is both orientation-dependent and limited by the saturation magnetic field of the phases involved. Achieving large MFIS through variant rearrangement therefore requires high-quality single crystals [11,14].

In contrast, the strain generated by MFIPT in these conventional FSMAs is generally low [15,16]. This is because the change in magnetization upon phase transformation from austenite (ferromagnetic/paramagnetic) to martensite (ferromagnetic/paramagnetic) is generally quite small [17,18]. Consequently, the shift in characteristic martensitic transformation temperatures with an applied magnetic field is also small, typically around 1 K/T. This necessitates extremely large magnetic fields to induce strain through MFIPT, which could be another reason for the low output stress observed [18]. Thus, for conventional FSMAs, a bulk single crystal is the primary requirement to attain large MFIS though martensite variant rearrangement. However, the technological challenges involved in producing high-quality bulk single crystals of specific compositions, combined with the high cost and low actuation stress, limit their applicability as practical actuator materials.

On the other hand, Ni-Mn-Sn/In based FSMAs show MFIS by virtue of MFIPT mechanism [5,7]. The magnetic field in these systems plays the same role as stress or temperature for conventional thermo-elastic shape memory alloys [19]. The key feature of these alloys is the significant change in magnetization ($\Delta M$) during the martensitic phase transformation from a ferromagnetic austenite phase to an anti-ferromagnetic/paramagnetic martensite phase [5,7,20,21]. This leads to a large shift in the martensitic transformation temperature with magnetic field (around 10-12 K/T) and consequently, a high MFIS under relatively low applied fields and results in a substantial actuation stresses (100 MPa) [5,22]. In this process, Zeeman energy plays a dominant role, overshadowing magneto-crystalline anisotropy energy (MCAE) [20,23]. It is worthwhile to mention here that the Zeeman energy depends both on the material's magnetization and the external magnetic field strength. Furthermore, the magnetization of the martensite and austenite phases are different, and this inherent magnetization also depends on magneto-crystalline anisotropy. Thus, Zeeman energy also inherently depends on magneto-crystalline anisotropy. Hence in single crystals, magnetization and hence Zeeman energy depends on the orientation of the crystal [24]. On the other hand, in polycrystalline materials this is averaged out and there is no orientation dependence of Zeeman energy, making them good candidates for actuator applications [11]. However, the major challenge with Ni-Mn-Sn/In polycrystalline FSMAs is their low MFIS (termed as spontaneous strain for unstrained Ni-Mn-Sn/In based FSMAs) arising from MFIPT. Typical MFIS values for these polycrystalline alloys are under 0.2% [25,26]. This limited strain hinders their practical applications. Substituting Co into the ternary Ni-Mn-Sn/In alloys proves to be an effective solution, boosting the MFIS to around 0.6% [27] along with increased Zeeman energy [11]. Co addition strengthens the ferromagnetic interaction in the austenite phase, enhancing the Curie temperature ($T_C$) and lowering the martensitic transition temperature ($M_s$ is the starting temperature of martensite transition and $A_s$ is the same for austenite transition) [28,29]. This results in: (i) increased $\Delta M$ and (ii) reduced entropy change ($\Delta S$) upon martensitic transformation [28]. $\Delta S$ value is found to maintain inverse relationship with ($T_C$ - $M_s$) [30] or ($T_C$ - $A_s$) [31]. In this context, it's important to note that a large $\Delta M$ (leading to high Zeeman energy, $\mu_0 \Delta M.H$; $\mu_0$ is permeability of free space and $H$ the applied magnetic field) and small $\Delta S$ are the key parameters for achieving high MFIS in Ni-Mn-Sn/In alloys with lower magnetic fields. This reduces the driving force (magnetic field) needed for MFIPT [32]. Additionally, low hysteresis and a martensitic transformation temperature close to room temperature further enhance the alloy's performance [33,34]. Therefore, Co-doped Ni-Co-Mn-Sn/In quaternary FSMAs hold significant promise for achieving high MFIS at low magnetic fields. Moreover, by tuning the alloy composition, their operational temperature range can be adjusted. Furthermore, incorporating pre-strain in the martensite phase can further enhance MFIS in these alloys. The ability to recover the pre-strain as MFIS with relatively low magnetic field will ultimately determine their effectiveness as actuator materials.

In a recent study on Ni-Mn-Sn/In based FSMAs, a large one-way MFIS with nearly full recovery (96.7%) for 3% pre-deformed $Ni_{45}Co_5Mn_{36.7}In_{13.3}$ single crystal for an applied magnetic field of 8 T has been reported by Kainuma et al.[5]. Following this work, a two-way shape memory effect (TWSME) has also been reported in 1.3% pre-strained polycrystalline $Ni_{43}Co_7Mn_{39}Sn_{11}$ alloy by Kainuma et al. [7]. In contrast to the single crystal result, polycrystalline $Ni_{43}Co_7Mn_{39}Sn_{11}$ alloy is seen to register a 77% recovery of the 1.3% pre-deformation strain for an applied magnetic field of 7 T and a TWSME has been observed between a strain value of 0.7% to 1% on field loading and unloading cycles [7]. A similar work by Ito et al. has shown only 0.56% recovery in 3.1% pre-strained spark plasma sintered polycrystalline $Ni_{43}Co_7Mn_{39}Sn_{11}$ alloy for an applied magnetic field of 8 T but the material ductility is improved significantly compared to the conventional induction-melted alloy [35]. Further, new materials exhibiting large spontaneous strain are also being explored in the present context [25–27,36]. For example, a large spontaneous strain of 1.7% was reported for polycrystalline $Ni_{50}Mn_{36}In_8Sb_6$ FSMA by Yu et al.[37]. In addition to these, researchers are actively pursuing the design of FSMAs that require lower magnetic fields for complete MFIPT through selection of suitable alloy composition, incorporation of precipitates in the matrix and thermo/ magneto-mechanical treatments.

In view of these, this study systematically investigates the influence of magnetic fields on the martensitic transformation temperature, magnetization, and shape memory properties of bulk polycrystalline $Ni_{37}Co_{11}Mn_{43}Sn_9$ alloy. We explore the possibility of achieving large MFIS by incorporating pre-strain in the martensite phase at room temperature. We further delve into the microstructural changes underlying the substantial recoverable MFIS and illustrate the mechanism involved.

## 2. Experimental procedure

The alloy button with a nominal composition of $Ni_{37}Co_{11}Mn_{43}Sn_9$ was prepared by vacuum arc melting of high purity (99.99 %) elements in appropriate proportions. Homogeneity was ensured by re-melting the alloy multiple times by flipping it upside down before each melting. The button was sealed in a quartz ampoule filled with helium gas, solutionized at 1123 K for 24 h, and quenched in ice water bath. Detailed characterizations of these alloys were carried out using scanning and transmission electron microscopy (SEM & TEM), x-ray diffraction (XRD), Differential Scanning Calorimetry (DSC), dc magnetization and strain gauge measurements. Samples for metallography were etched using an aqueous solution of $FeCl_3$ in HCl. XRD experiments were carried out for both bulk and powder samples using a Cu Kα radiation. TEM was performed by employing a JEOL 2000FX microscope operating at 200 kV. Specimens for TEM were prepared by slicing discs from an electro-discharge machined cylindrical rod of 3 mm diameter, followed by grinding and jet polishing with a Struers Tenupol-5 at 233 K, using a 10 vol.% perchloric acid in methanol electrolyte. Isochronous DSC experiments were performed using a Mettler-Toledo calorimeter at a rate of 10 K/min in argon atmosphere. Magnetometry was carried out using a commercial 9 Tesla vibrating sample magnetometer (VSM) based on a Physical Property Measurement System (PPMS) (Quantum Design Inc., USA). Strain was measured by using the strain gauge technique as a function of temperature and magnetic field in a home-made variable temperature insert working in a commercial magnet-cryostat system (American Magnetics Inc., USA). A 5 mm x 5mm x 10 mm sized sample was used for strain measurements. Cu has been utilized as the reference material for the differential measurement of relative length change ($\Delta l/l$). Literature reported data for Cu [38] were used to calculate the temperature and magnetic field dependence of $(\Delta l/l)_H$ for the $Ni_{37}Co_{11}Mn_{43}Sn_9$ alloy. The temperature dependence of strain is evaluated as the relative length change *($\Delta l/l$)* compared to the sample's length at 293 K. Similarly, the isothermal magnetic field dependence is measured as $(\Delta l/l)_H$ relative to the zero-field length.

## 3. Results and discussion

*3.1. Microstructural investigation:*

Microstructure of the $Ni_{37}Co_{11}Mn_{43}Sn_9$ alloy has been studied at room temperature in the as-solutionized condition and a two-phase microstructure is observed as shown in representative SEM image (Fig.1). Typical grain size is found to be around 400 μm. The microstructure comprises mainly of martensite matrix and some fraction of *γ*-phase at the grain boundaries. This is expected as the martensite start temperature ($M_s$) for the alloy is close to room temperature (presented in details in section 3.2). The inset of Fig. 1 shows the representative SEM image for the magnified portion of the twinned martensite matrix and *γ*-phases at the grain boundaries. The overall composition of the bulk alloy as determined from X-ray Fluorescence (XRF) analysis is in close agreement with the nominal composition, presented in Table 1. The chemical composition for the matrix and γ-phase, as determined by EPMA, is listed in Table 2. The errors in the composition analysis correspond to one standard deviation. Thus, from the composition analysis, it can be inferred that the γ-phase is rich with Co and Mn while it is lean with Ni and Sn. Valence electron concentration per atom ratio ($e/a$) is also computed for this alloy, in terms of the concentration weighted sum of the number of $3d$ and $4s$ electrons of Ni, Mn, Co and the number of $5s$ and $5p$ electrons of Sn, as can be seen in Table 1.

*3.1. Thermal analysis:*

Figure 2 shows the DSC plots of the $Ni_{37}Co_{11}Mn_{43}Sn_9$ alloy in its as-solutionized condition. The plot clearly reveals evidence of a reversible structural martensitic transformation for this alloy. The ferromagnetic to paramagnetic transition, visible in the DSC scans, is characterized by a change in the baseline slope for the austenite phase. This transition is represented by the dashed line in the Fig.2, indicating the Curie temperature ($T_C$). Characteristic martensitic transformation temperatures, including martensite start ($M_s$), martensite finish ($M_f$), austenite start ($A_s$), and austenite finish ($A_f$) temperatures, along with $T_C$ and transformation enthalpies ($\Delta H$), are listed in Table 3. The entropy changes associated with the transformation ($\Delta S$) for both cooling and heating processes are calculated by dividing the average of the forward and reverse transformation enthalpies, respectively, by the reference temperature $T_0$. This value is tabulated in Table 3. $T_0$ is defined as the transformation temperature at which the parent phase and the product martensite phase have the same Gibbs free energy, given by ($A_f + M_s$)/2 [31,39].

*3.3. XRD and TEM analyses:*

Figure 3 displays XRD pattern of $Ni_{37}Co_{11}Mn_{43}Sn_9$ at room temperature (300 K). The full profile of the data is fitted using the Le Bail method, which revealed a mixture of three phases:

austenite with the $L2_1$ crystal structure (*Fm3m*), face-centered cubic (fcc) γ-phase (*Fm-3m*), and a six-layer modulated monoclinic (6M) martensite with the *P21* space group. Inset of Fig. 3 shows magnified portion of the data between 2θ values of 38° to 48°, which corresponds to a good fit. The corresponding cell parameters are: (i) $a$ = 5.9281(3) nm for the $L2_1$ structure, (ii) $a$ = 3.8995(8) nm for the γ-phase and (iii) $a$ = 4.4779(2) nm, $b$ = 5.4787(8) nm, $c$ = 25.617(2) nm, $β$ = 94.7921$^0$(2) for the 6M martensite. These results agree with the XRD study by Umetsu et al. for the same composition, which also identified $L2_1$ austenite and both 6M and 14M martensites at room temperature [40]. However, they did not report the presence of an fcc γ-phase. Figure 4 shows the bright-field TEM micrograph of $Ni_{37}Co_{11}Mn_{43}Sn_9$ at room temperature that depicts fine martensite plates. Correspondingly, Selected Area Electron Diffraction (SAED) pattern (inset of Fig. 4) clearly shows evidence for 6-layer modulation in terms of the satellite spots. This TEM analysis confirms the XRD findings regarding the martensite configuration of the alloy at room temperature.

*3.4. Magnetization measurement:*

Dc-magnetization measurements on $Ni_{37}Co_{11}Mn_{43}Sn_9$ alloy were carried out in presence of varying magnetic field strengths between 0.05 to 9 T, as shown in Fig. 5(a). Initially the magnetization was recorded by following the field cooled cooling (FCC) convention by applying the magnetic field at 400 K and cooling it to 5 K. Subsequently, the alloy was heated back to 390 K in presence of the same field strength and data were collected in field cooled warming (FCW) mode. This protocol was maintained for all the field strengths as displayed in Fig. 5(a). Upon cooling from 400 K in the FCC protocol, the magnetization drops suddenly around 292.7 K (for $H$ = 0.05 T), which corresponds to the martensitic transformation (austenite to martensite) of the alloy whereas the sharp rise of magnetization around 300.6 K (for $H$ = 0.05 T) is observed upon heating under FCW protocol, which corresponds to the reverse martensitic (martensite to austenite) transformation. The thermal hysteresis observed between the FCC and FCW curves is a characteristic feature of a first order martensitic transformation. The increase in applied magnetic field causes the characteristic martensitic transformation temperatures to decrease. The change in temperature ($\Delta T$) for different magnetic fields is always noted with respect to the DSC obtained zero field data. For example, $M_s$ was found to reduce to 207.7 K on application of 9 T field from 295.6 K ( for zero field data, see DSC plot, Fig.2). Figure 5(b) depicts the variation of characteristic temperatures with magnetic field. The linear fitting of the data provides the rate of decrease of these characteristic martensitic transition temperatures. For example, $M_s$ decreases at a rate of 9 K/T. The decrease of these characteristic martensitic transition temperatures with magnetic field is attributed to the magnetic field induced stabilization of austenite phase.

Figure 5(c) shows the variation of the magnetization difference ($\Delta M = M_M - M_A$; $M_M$ and $M_A$ are the magnetization of the martensite and austenite phases, respectively) between the martensite and austenite phases during FCC and FCW protocols under different magnetic fields. The inset in the top left corner of Fig. 5(a) illustrates the method for calculating $\Delta M$, using values right below the martensite finish temperature ($M_f$) and right above the austenite start temperature ($A_s$). Initially, the magnitude of $\Delta M$ increases with increasing magnetic field and attains a maximum value of 93 Am²/kg at 2 T. Thereafter, it starts to decrease to some extent. This increase is attributed to field-induced austenite stabilization. Under such conditions, the final low-temperature matrix consists of transformed martensite with low saturation magnetization and some untransformed/arrested austenite with high saturation magnetization. Higher applied fields lead to increased stabilization of austenite, consequently raising its volume fraction within the martensite microstructure. This untransformed austenite volume fraction ultimately governs the $\Delta M$ variation. Similar variation of $\Delta M$ has been observed by Karaca et al. in $Ni_{45}Mn_{36.5}Co_5In_{13.5}$ single crystal [11] and Ito et al. in $Ni_{45}Mn_{36.7}Co_5In_{13.3}$ polycrystalline alloy [41]. They ascribed this variation to magnetic field-induced austenite retention or kinetic arrest of the martensitic transformation, evidenced by abnormal entropy changes and extremely low mobility of phase interfaces at low temperatures.. Conversely, a similar work on $Ni_{50}Mn_{34}In_{16}$ polycrystalline alloy by Krenke et al. [26] did not observe similar variation of $\Delta M$ with applied magnetic field. In their case, $\Delta M$ continuously increased with increasing field for the polycrystalline $Ni_{50.3}Mn_{33.8}In_{15.9}$ alloy, reaching saturation at around 1 T. This suggests that the addition of Co in NiMnSn/In alloys promotes kinetic arrest of the austenite to martensite transition. Experimental evidence for this arrest has been reported in Ni-Co-Mn-Sn/In-based FSMAs using temperature-dependent magnetization [41] and neutron diffraction [42]. At this point, it is worth noting that the hysteresis ($\Delta T_h$) across the transformation, which is given by ($A_f - M_s$), increases with increase in magnetic field and the same has been presented in Fig. 5(c) as well.

Figure 6(a), on the other hand, presents the first quadrant isothermal magnetization measurements under varying magnetic field (0 to + 9 T)across three distinct temperature regions, viz. completely inside martensite region (5 K and 150 K), martensitic transformation region (250 K -300 K) and austenite region (T > 325 K). These measurements were conducted in zero field cooling (ZFC) mode, i.e. the sample was first cooled from 399 K down to 5 K in zero field and the sample was then warmed up to the desired temperature before subjecting it to a field excursion. The first temperature region lies significantly below the martensitic transformation. Here, $M$ vs $H$ curves at 5 K and 150 K represent the magnetization behaviour for the alloy. These curves exhibit ferromagnetic state of martensite. The applied field strength in this region is insufficient to cause the structural transformation to austenite phase. The austenite region encompasses temperatures above $A_f$ temperature. The $M$ vs $H$ curves at 325 K, 360 K and 399 K showcase the behavior in this region. These curves exhibit a typical response of a soft ferromagnetic material where the magnetization increases quickly and saturates at low fields (< 0.45 T) with negligible hysteresis. The martensite transformation region spans a temperature range in the vicinity of the characteristic martensitc transformation temperatures but below $A_f$ temperature. M vs H curves between 250 K and 305 K correspond to this region. Here, we observe a metamagnetic-like transition, where an applied magnetic field triggers a transformation from the low-moment martensitic phase to the higher-moment austenitic phase. Intriguingly, the $M$ vs $H$ curves at 295 K, 300 K, and 305 K exhibit a magnetic field-induced martensite-to-austenite transformation. However, the reverse transformation (austenite to martensite) fails to occur upon field removal. This is related to the fact that at these temperatures $M_S$ line (see Fig. 5(b) ) is hardly crossed while decreasing field to zero, resulting in zero or negligible martensite formation. The field analogues of martensite start ($H_{Ms}$), martensite finish ($H_{Mf}$), austenite start ($H_{As}$) and austenite finish ($H_{Af}$) are extracted from the metamagnetic $M$ vs $H$ curves. Figure 6(a) exemplifies this extraction for the 250 K curve. Figure 6(b) depicts the variation of the as derived critical transformation fields with temperature. We observe a decrease in critical field values with increasing temperature. This is because the critical field required to drive the martensitic transformation reduces as the measurement temperature approaches the martensitic transformation temperature (for example, $A_s$ temperature). By performing a linear fit on these critical field variations, we can calculate the rate of their decline with increasing temperature. For example, $H_{Ms}$ decreases at a rate of 0.12 T/K, while for $H_{Af}$ the value is 0.11 T/K.

It is even possible to estimate the transformation entropy (ΔS) change upon martensitic transformation from the magnetization data, utilizing the Clausius–Clapeyron (CC) relation in the magnetic phase diagram. CC relation is defined as [5,43]:

$$\Delta S = \Delta M \left(\frac{dH_{Ms}}{dT}\right) \text{ and } \Delta S = \left(\frac{\Delta M}{\Delta T}\right)\Delta H \qquad (1)$$

Where ΔM is the change in magnetization upon martensitic transformation. The ΔM value at $M_s$=275 K (as obtained from Fig. 6(a)) is found to be 98.79 Am$^2$/Kg while $dH_{Ms}/dT$ is found to be 0.12 T/K. Thus, the ΔS value as predicted by CC equation (i.e. Eq.(1)) is 11.85 J /Kg-K. On the other hand, the ΔS value of the Ni$_{37}$Co$_{11}$Mn$_{43}$Sn$_9$ alloy is calculated from the enthalpy data obtained from the DSC (under zero field) as 11.51 J /Kg-K. Thus, transformation entropy change upon MT as estimated from CC equation is in close agreement with DSC result. Similar result is also available in literature [44]. The estimated values of ΔS obtained from Eq. (1) using ΔT and ΔM for the change in magnetic field (ΔH) is plotted as an inset in Fig. 5(c). Interestingly, the ΔS value decreases with increasing magnetic field above 2 T. This can be attributed to the arrested ferromagnetic austenite phase, which leads to a lower ΔM across the transition.

*3.5. Thermo-magnetic strain:* Figure 7 shows the temperature dependence of spontaneous transformation strain (*Δl/l*, defined as $(l(T)-l_0)/l_0$; $l_0$ being the length of the specimen at 293 K [25]) in zero magnetic field for: (i) unstrained, and (ii) 1.3% pre-strained Ni$_{37}$Co$_{11}$Mn$_{43}$Sn$_9$ specimens, measured during a cooling-heating cycle. Pre-strained specimen was prepared by compressing the polycrystalline alloy specimen (5 mm x 5 mm x 10 mm) along long axis at room temperature (in the martensite state as testing temperature is well below $A_s$ temperature of 318.6 K) and the residual strain was 1.3 %. Now, the temperature induced strain evolution curves for both the samples show similar behavior. Starting from well above the $A_f$ temperature (austenite phase), *Δl/l* initially decreases smoothly with decreasing temperature. Then, the curves show a sharp drop near 290-300 K, followed by another smooth decrease. Upon heating from well below the $M_f$ temperature, a very similar behavior is observed, but with a distinct thermal hysteresis between 280 and 320 K. The temperatures at which the strain undergoes a significant change correspond to the forward (cooling curves) and reverse martensitic (heating curves) transitions. The characteristic martensitic transition temperatures ($M_s$, $M_f$, $A_s$ and $A_f$) measured for the unstrained and pre-strained specimens are presented in Table 4. These values demonstrate close agreement with those obtained through DSC analysis for unstrained polycrystalline $Ni_{37}Co_{11}Mn_{43}Sn_9$ alloy. Notably, pre-straining the specimen by 1.3% leads to an almost 5 K increase in the characteristic martensitic transition temperatures and a 1.9 K decrease in thermal hysteresis.

Similarly, the isothermal magnetic field induced strain $(\Delta l/l)_H$ is defined as $(l(H)-l_0)/l_0$ [26], where $l_0$ is the length of the specimen in the absence of field and $l(H)$ the length in the presence of field. Figure 8(a) present the $(\Delta l/l)_H$ vs $H$ curves for both unstrained and 1.3 % pre-strained $Ni_{37}Co_{11}Mn_{43}Sn_9$ alloys. Measurement was carried out by increasing the magnetic field to 5 T and subsequently reducing it back to zero. Magnetic field was applied parallel to the long axis (10 mm) of the specimen (or along the compression axis for pre-strained specimen) at room temperature. The unstrained sample begins to expand at an applied magnetic field around 1.2 T, and then the strain sharply increases to 0.25 % at 2.1 T. However, no reversible length change occurs during magnetic field removal. This is because the applied field induces a reverse martensitic transformation, but removing the field fails to initiate transformation due to the field-induced stabilization of the higher-magnetic-moment austenite phase. This effect is clearly visible in the inset of Fig. 8(a), which shows the isothermal $M$ vs $H$ plot at 300 K. Here, the magnetic field-induced transformation from martensite to austenite starts at 1.2 T and completes at 2.8 T. These values closely match the field values observed in the MFIS measurement (main panel of Fig. 8(a)). This one-to-one correlation between the isothermal $M$ vs $H$ plot and the $(\Delta l/l)_H$ vs $H$ plot for the unstrained specimen confirms that the MFIS is primarily due to the magnetic field-induced martensite-to-austenite transformation. The small observed MFIS, or spontaneous strain, can be attributed to the associated volume change from martensite to austenite [32,47]. In contrast, no transformation is observed upon removal of the field due to the stabilization of the austenite phase. This finding agrees with previous reports, which have shown magnetic field-induced stabilization of the austenite phase and even kinetic arrest effects in similar Ni-Mn-Sn/In-based alloys [42,43,48]. Therefore, the one-way magnetic field-induced shape memory effect observed in the unstrained $Ni_{37}Co_{11}Mn_{43}Sn_9$ alloy at room temperature aligns well with the isothermal magnetization behavior of the alloy.

In comparison, unlike the unstrained sample, the pre-strained specimen demonstrates a significantly larger MFIS. As depicted in Fig. 8(a), the MFIS starts rising rapidly at 2.8 T and plateaus at 1.55 % at 4.45 T. Notably, this recovered strain (1.55 %) exceeds the applied pre-strain of 1.3%, suggesting that it comprises both spontaneous strain (0.25%, due to crystal dimension change from martensite to austenite) and the pre-deformation itself. This finding highlights the potential for achieving remarkable MFIS in Ni-Mn-In/Sn-based FSMAs by combining high spontaneous strain with substantial pre-deformation in the martensite phase. The pre-strain limit of an alloy, dictated by its inherent ductility, is a major challenge for Ni-Mn-In/Sn-based FSMAs, known for their extreme brittleness. Fortunately, incorporating the γ-phase (does not take part in martensitic transformation) into the matrix through suitable alloy composition and suitable thermo-mechanical processing can enhance the ductility of these materials [35,49,50]. Furthermore, this study surpasses previous results by achieving a higher recovery percentage (100% of the applied 1.3% pre-strain) with a lower magnetic field (4.45 T). For example, Ref. [7] reported a 77% recovery of 1.3% pre-strain in $Ni_{43}Co_7Mn_{39}Sn_{11}$ using a 7 T field, and Ref. [35] reported an 18% recovery of 3.1% pre-strain in spark plasma sintered $Ni_{43}Co_7Mn_{39}Sn_{11}$ with an 8 T field. The lower critical magnetic field of 4.45 T for complete MFIPT in the current 1.3% pre-strained polycrystalline $Ni_{37}Co_{11}Mn_{43}Sn_9$ alloy specimen is attributed to the combined effect of a lower transformation hysteresis, a smaller transformation interval ($\Delta T_i = (A_f - A_s)$), and a larger $\Delta M/\Delta S$ value (critical field is approximated as, $(\mu_0 \Delta M)_{min} \approx (\Delta T_h + \Delta T_i) / (\Delta M/\Delta S)$) [11,38]. Introduction of 1.3% pre-strain reduces $\Delta T_h$ and $\Delta T_i$ by 1.4 K and 1.9 K, respectively, as presented in Table 4. Upon field removal, the pre-strained specimen exhibits a significant length change of around 0.55% (strain value reduces to 1.0 % from 1.55%). Figure 8(b) demonstrates the reversible magnetic field-induced TWSME between strain values of 1% and 1.55% for three consecutive cycles on application of magnetic field in the range of 0 - 4.45 T. This remarkable recovery of reversible two-way shape, achieved with minimal magnetic field application, significantly surpasses previously reported values. For instance, Kainuma et al. [7] observed a 0.3% reversible strain recovery (between 0.7% and 1%) in a 1.3% pre-strained polycrystalline $Ni_{43}Co_7Mn_{39}Sn_{11}$ alloy subjected to a 7 T magnetic field cycle. In contrast, Ito et al. reported only a faint reversibility (~0.04% strain) for a spark plasma sintered 3.1% pre-strained polycrystalline $Ni_{43}Co_7Mn_{39}Sn_{11}$ alloy at 310 K when undergoing an 8 T magnetic field cycle [35].

Figure 9 illustrates the mechanisms behind magnetic field-induced strain recovery in both unstrained and 1.3% pre-strained specimens. The unstrained sample exhibits a small MFIS value of 0.25% due to MFIPT from a multi-variant self-accommodating martensitic structure to the parent austenite phase. However, a large MFIS requires a MFIPT from single variant martensite to austenite [11]. In the current unstrained alloy, the magnetic field does not favor any specific martensite variant due to their weak magnetic moment and small magneto-crystalline anisotropy. Consequently, the MFIS from MFIPT remains limited here. Now, pre-straining the alloy causes certain martensitic variants, which are in favorable orientation with respect to the stress, to grow at an expense of others through twin boundary movements. Therefore, pre-strained specimen resides in stress-favorable variant martensitic configuration and application of magnetic field causes a transformation from the stress-favorable martensite variant configuration to parent austenite. In this process, a large amount of strain is recovered which consists of pre-strain which is being stored in the stress-favorable variant configuration and strain due to crystallographic dimension change. In this case, the role of pre-strain is to bias a specific martensite variant while magnetic field (as it cannot directly bias variants due to small magneto-crystalline anisotropy and low magnetization of the martesite phase of the Ni-Mn-Sn based FSMAs) plays the role of temperature to induce reverse martensitic phase transformation. In contrast, for pre-strained specimen, field unloading triggers the magnetic field-induced transformed austenite back to martensite, resulting in a length change of 0.55% (from 1.55% strain to 1.0% strain). However, this transformation is only partial, as the magnetic field is known to stabilize the austenite phase. Xuan et al. have experimentally observed such partial transformation in Ni-Co-Mn-Al based FSMA systems using magnetic force microscopy [49]. In contrast, the transformation is ceased for unstrained specimen due to magnetic field induced complete stabilization of austenite. Notably, incorporating 1.3% pre-strain triggers a transition from one-way SME to TWSME. The observed shift to TWSME likely stems from the combined effect of smaller $\Delta T_h$ (1.4 K), narrower $\Delta T_i$ (1.9 K) and increase in $M_s$ temperature (5.4 K), as displayed in Table 4. Increase in $M_s$ temperature by 5.4 K causes measurement temperature (300 K) to overlap with $M_s$. This reduces the driving force required for austenite to martensite transformation upon field removal. These combined effects, aligned with the predicted decrease in critical field for MFIPT, ultimately promoting TWSME. Finally, the pre-strained specimen exhibits fully reversible TWSME between strain values of 1.0% to 1.55 %, tested for three consecutive cycles, indicating stabilization of the stress-induced martensite variant configuration for zero field conditions. The underlying mechanism for this TWSME can be attributed to the creation and pinning of preferred nucleation sites for specific martensite variants, facilitated by magnetic field-training [50]. Furthermore, constraint stress from surrounding grains can also promote this creation and pinning of preferred nucleation sites for specific martensite variants in the current polycrystalline $Ni_{37}Co_{11}Mn_{43}Sn_9$ alloy [7,51].

## 4. CONCLUSIONS

Microstructural investigation for bulk polycrystalline $Ni_{37}Co_{11}Mn_{43}Sn_9$ alloy reveals that room temperature microstructure comprises mixture of martensite matrix (6M), austenite with $L2_1$ and FCC γ-phase at grain boundaries. The alloy shows martensitic transformation in close proximity of room temperature with a low associated entropy change of 11.51 J/kg-K. Thermo-magnetization curves under constant magnetic field demonstrate magnetic field induced reversible martensitic transformation between ferromagnetic austenite with higher saturation magnetization and martensite with lower saturation magnetization. The effect of magnetic field on martensitic transformation temperature is manifested with a rapid decrease of characteristic temperatures (at a rate of 9 K/T for $M_s$, as an example) with increase in magnetic field which envisages large MFIS. MFIS study for the unstrained virgin specimen shows one way shape memory effect with 0.25 % spontaneous strain with application of 2.1 T magnetic field. In contrast, MFIS study on 1.3% pre-strained specimen registers TWSME with spontaneous length change of 0.55% (between strain values 1.55% to 1%), tested for 3 consecutive magnetic field cycles. This result suggests that for pre-strained specimen: (i) spontaneous strain (0.25 %) and pre-strain (1.3 %) together (a total of 1.55 % strain) is recovered upon MT through MFIRT; (ii) full recovery (100%) of 1.3% pre strain with application of just 4.45 T magnetic field. This 100 % strain recovery is found to be 23% higher than the existing highest reported value of 77 % in literature for polycrystalline FSMAs. Similarly, the reversible recovered strain of 0.55% during TWSME is nearly double the highest reported value for these materials (0.3%). Moreover, the requirement of magnetic field for this complete shape recovery appears to be only 4.5 T while earlier reports with polycrystalline FSMAs demand much larger field (> 5 T). The combination of near room-temperature martensitic transformation, 100% pre-strain recovery, and a relatively low required magnetic field (4.45 T) make the polycrystalline $Ni_{43}Co_7Mn_{39}Sn_{11}$ alloy particularly promising for practical actuator applications. Furthermore, this work demonstrates that pre-straining materials with high spontaneous strain can lead to remarkably high MFIS values. This finding opens new avenues for designing actuator materials with significantly enhanced shape memory response, offering potential in various applications.

**Acknowledgements:**

Authors acknowledge the contributions of Dr. Aniruddha Biswas, Dr. Arijit Laik and Mr. Santosh Yadav from Materials Science Division, Bhabha Atomic Research Centre, Mumbai in the work.

**Figures**

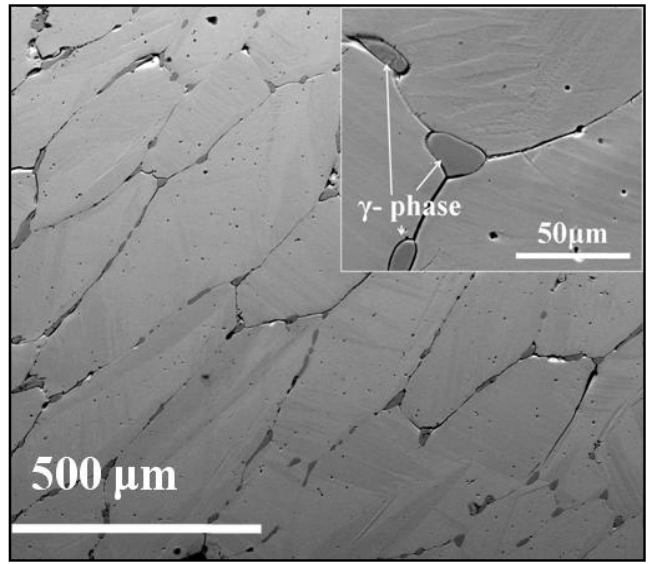

**Fig. 1.** Representative SEM image showing martensite matrix and γ-phase at boundaries for $Ni_{37}Co_{11}Mn_{43}Sn_9$ alloy. The inset shows magnified portion of twinned martensite (upper right area in the image).

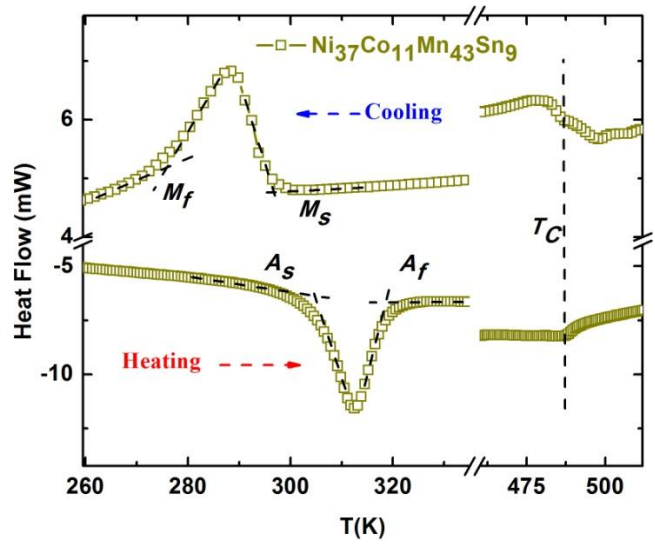

**Fig. 2.** DSC plot showing characteristic MT temperatures and Curie temperature for $Ni_{37}Co_{11}Mn_{43}Sn_9$ alloy.

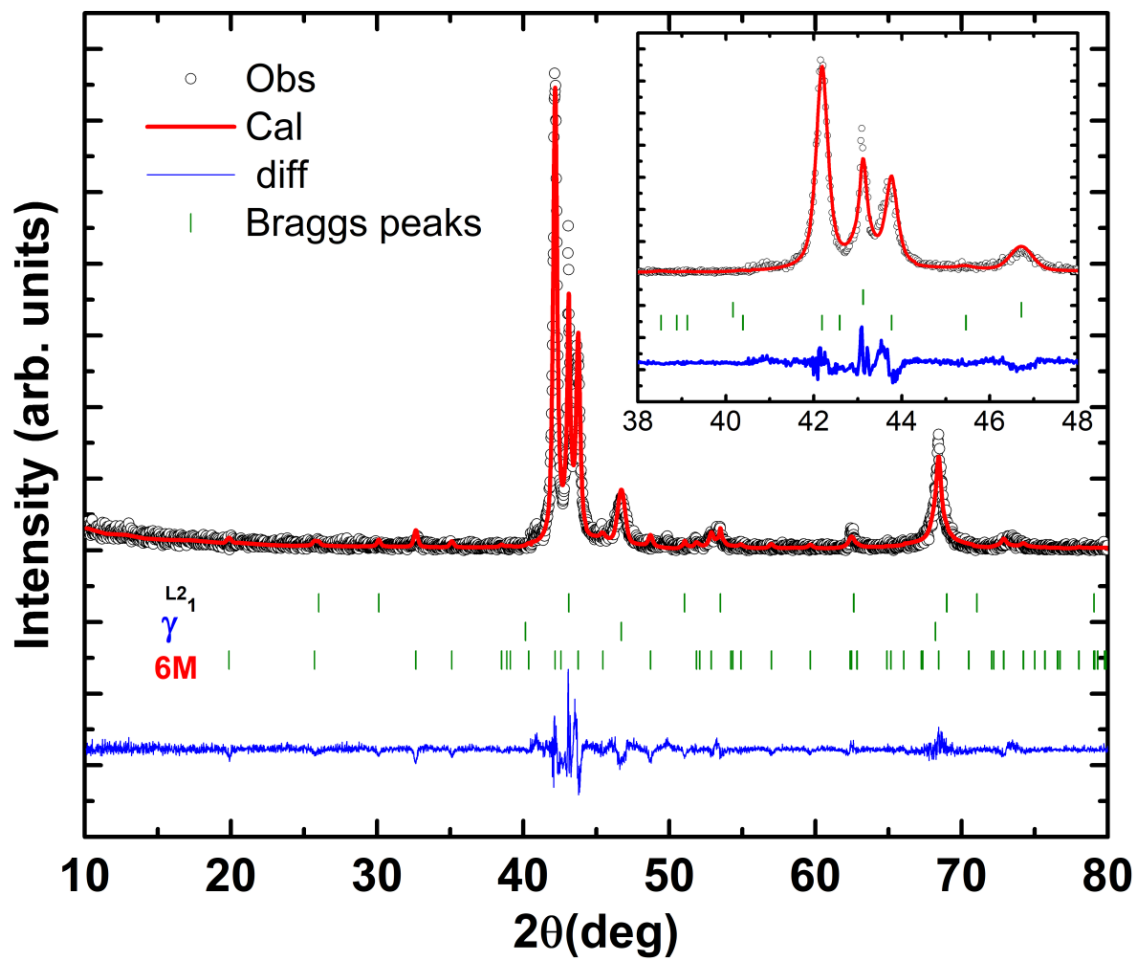

**Fig. 3.** Le-Bail fitted XRD pattern at RT for $Ni_{37}Co_{11}Mn_{43}Sn_9$ alloy. Black open circle shows the observed data, red line shows the calculated pattern, blue line depicts the difference pattern and green ticks represents the Bragg peaks. Inset shows magnified portion (between 2θ values of 38° to 48°) of the data.

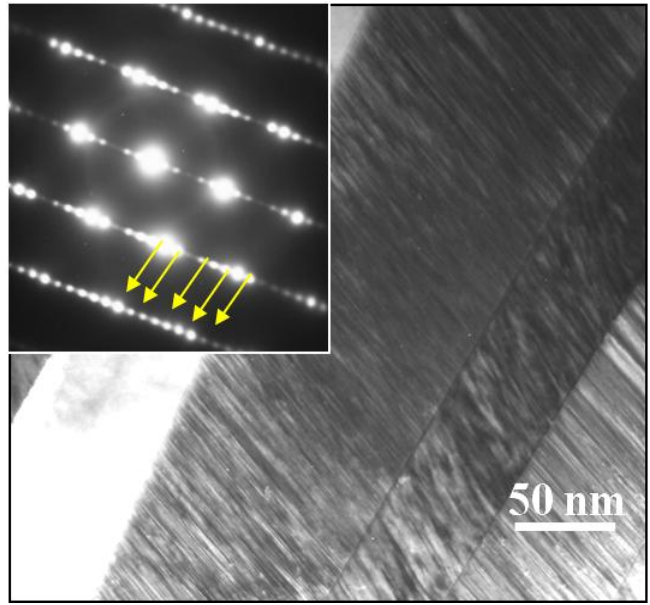

**Fig. 4.** Bright Field TEM image shows martensite plates for $Ni_{37}Co_{11}Mn_{43}Sn_9$ alloy while inset (upper left) shows satellite spots in SAED image as evidence of 6M.

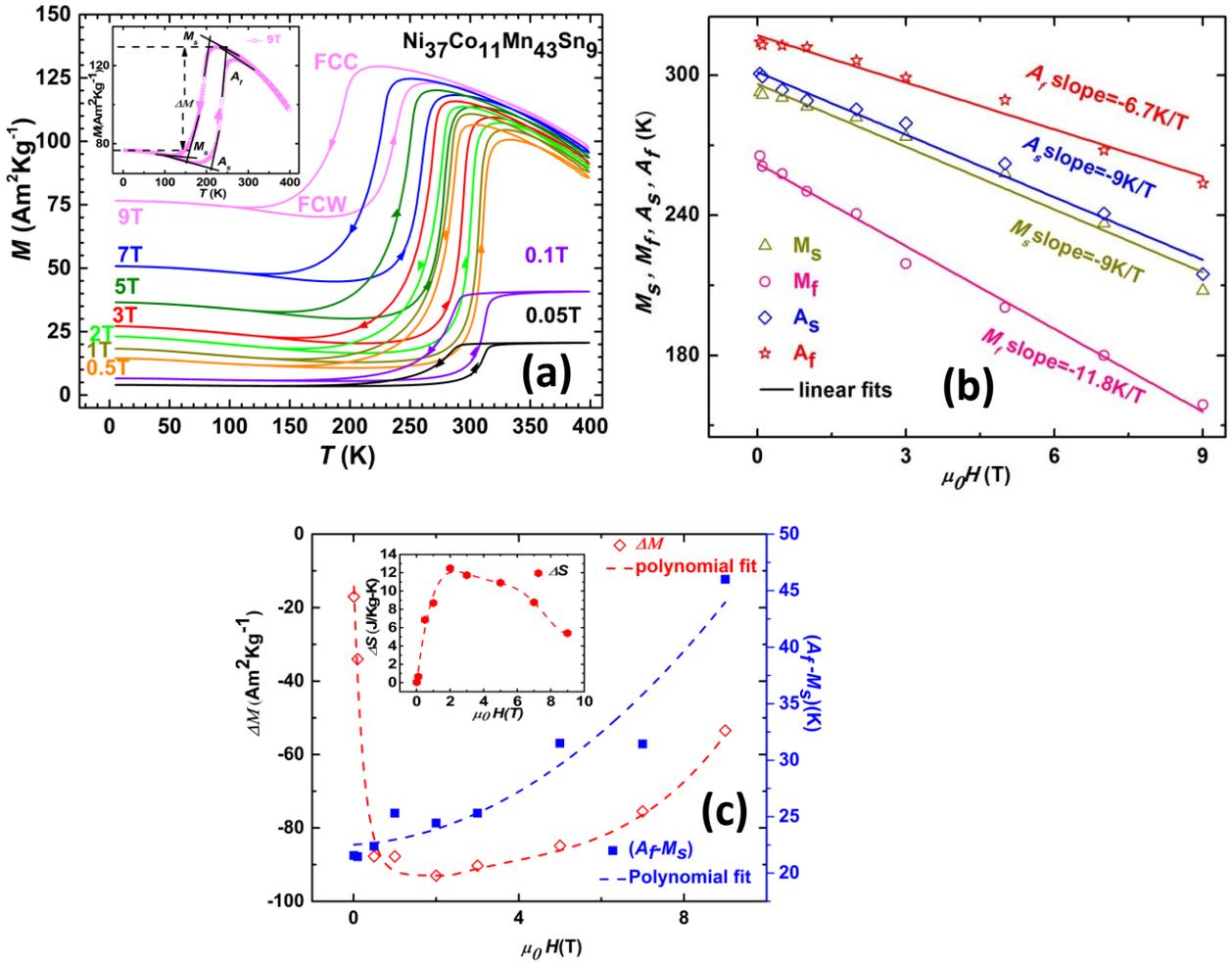

**Fig. 5.** For Ni$_{37}$Co$_{11}$Mn$_{43}$Sn$_9$ alloy: (a) Thermo-magnetization curves as a function of applied magnetic field; inset shows estimation of characteristic martensitic transformation temperatures and change in magnetization ($\Delta M$) upon martensitic transformation for an applied field of 9 T, shown as an example, (b) Variation of characteristic martensitic transformation temperatures as a function of applied magnetic field (c) Magnetization change and thermal hysteresis as a function of applied field while inset shows the associated entropy change upon martensitic transformation with applied field.

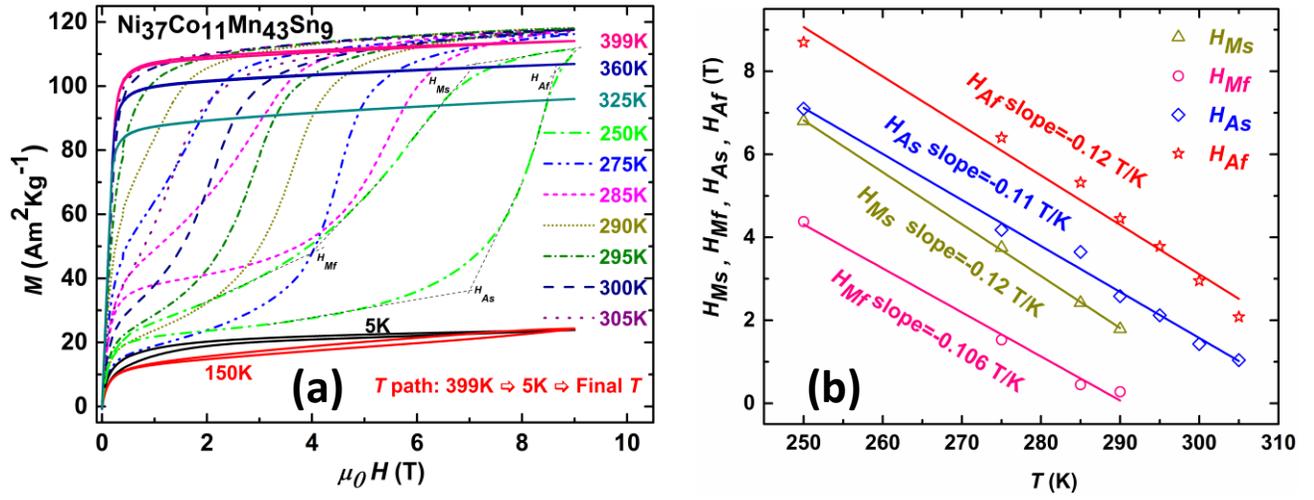

**Fig. 6.** For $Ni_{37}Co_{11}Mn_{43}Sn_9$ alloy: (a) Isothermal magnetization curves measured at temperatures from 5 K to 399 K, (b) Critical field-temperature phase diagram.

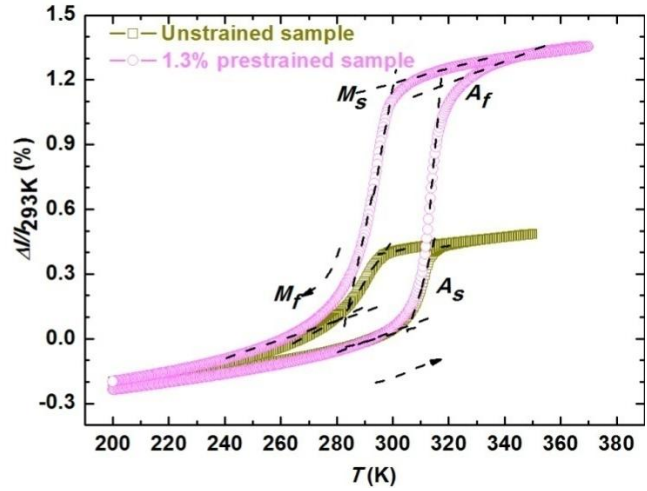

**Fig. 7.** Temperature dependence of strain for unstrained and 1.3 % pre-strained polycrystalline $Ni_{37}Co_{11}Mn_{43}Sn_9$ alloys under zero applied magnetic field.

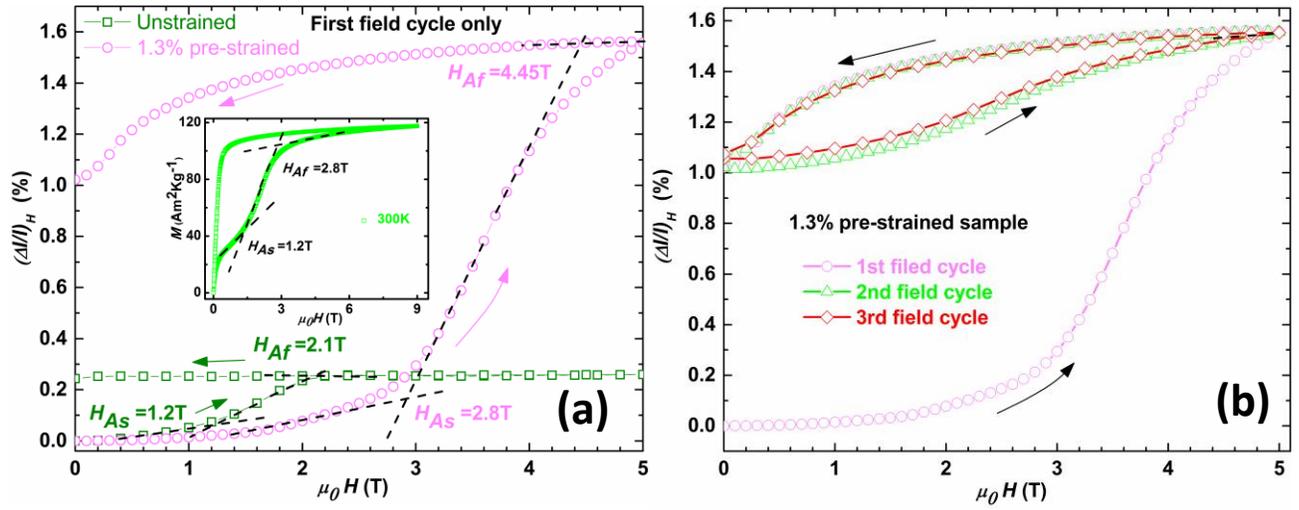

**Fig. 8.** (a) Isothermal magnetic field dependence of strain at 300 K for polycrystalline $Ni_{37}Co_{11}Mn_{43}Sn_9$ alloy for unstrained and 1.3% pre-strained specimen, shown for the first field cycle only; inset shows corresponding isothermal magnetization plot at 300 K, (b) TWSME for 1.3% pre-strained specimen shown for 3 consecutive field loading-unloading sequences.

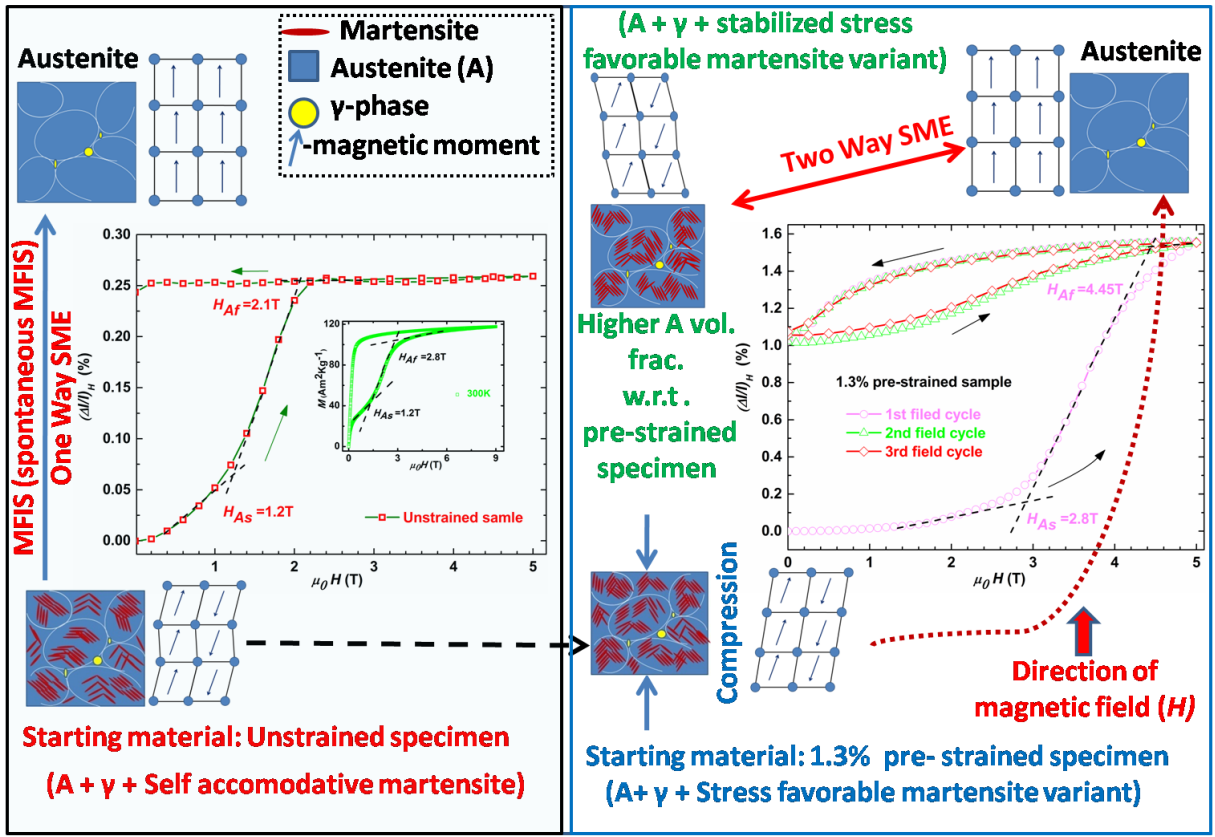

**Fig. 9.** Schematics for the associated mechanism of magnetic field induced strain recovery at 300 K for polycrystalline $Ni_{37}Co_{11}Mn_{43}Sn_9$ alloy for unstrained and 1.3% pre-strained specimen.

**Tables**

| Alloy | Ni (at.%) | Co (at.%) | Mn (at.%) | Sn (at.%) | *e/a* |
|---|---|---|---|---|---|
| $Ni_{37}Co_{11}Mn_{43}Sn_9$ | 36.1±0.8 | 10.9±0.4 | 43.9 ±0.4 | 9.1 ±0.2 | 8.03 |

**Table 1:** XRF results showing overall bulk chemical composition of $Ni_{37}Co_{11}Mn_{43}Sn_9$ alloy.

| Elements | Matrix (at.%) | | γ-phase (at.%) | |
|---|---|---|---|---|
| | mean | σ | mean | σ |
| Ni | 37.07 | 0.24933 | 32.28 | 0.3308 |
| Co | 10.27 | 0.1156 | 17.82 | 0.42903 |
| Mn | 43.73 | 0.28553 | 47.58 | 0.4327 |
| Sn | 8.93 | 0.11311 | 2.32 | 0.50747 |

**Table 2:** EPMA result shows composition of matrix and γ-phase for $Ni_{37}Co_{11}Mn_{43}Sn_9$ alloy.

| Alloy | $M_s$ (K) | $M_f$ (K) | $A_s$ (K) | $A_f$ (K) | $T_C$ (K) | $\Delta H^{M \to A}$ (J/gm) | $\Delta H^{A \to M}$ (J/gm) | $T_0$ (K) | $\Delta S$ (J/Kg-K) |
|---|---|---|---|---|---|---|---|---|---|
| $Ni_{37}Co_{11}Mn_{43}Sn_9$ | 295.6 | 275.5 | 304.5 | 318.6 | 487.0 | -3.85 | 3.21 | 307.1 | 11.51 |

**Table 3:** Characteristics martensitic transformation temperatures, enthalpy change and entropy change values for the polycrystalline $Ni_{37}Co_{11}Mn_{43}Sn_9$ alloy, obtained from DSC scans.

| Alloy | $M_s$ (K) | $M_f$ (K) | $A_s$ (K) | $A_f$ (K) | $\Delta T_i=(A_s-A_f)$ (K) | $\Delta T_h=(A_f-M_s)$ (K) |
|---|---|---|---|---|---|---|
| Unstrained $Ni_{37}Co_{11}Mn_{43}Sn_9$ | 294.6 | 270.7 | 304.4 | 314.5 | 10.1 | 19.9 |
| 1.3 % pre-strained $Ni_{37}Co_{11}Mn_{43}Sn_9$ | 300.0 | 284.2 | 309.3 | 318.0 | 8.7 | 18.0 |

**Table 4:** Change in characteristics martensitic transformation temperatures with incorporation of 1.3% pre-strain in polycrystalline $Ni_{37}Co_{11}Mn_{43}Sn_9$ alloy, along with hysteresis and transformation interval values, as obtained from temperature induced strain measurements.